\documentclass[12pt]{article}

\usepackage{amsmath,amssymb,amsthm}

\newcommand{\real}{\mathbb{R}}

\newcommand{\vj}{\boldsymbol{j}}
\newcommand{\vv}{\boldsymbol{v}}

\newcommand{\vs}{\boldsymbol{s}}

\newcommand{\vtri}{\boldsymbol{\vartriangle}}

\newcommand{\Gammav}{\varGamma}

\begin{document}

\title{\bf  
 Vortex Lattices \\
 in Quantum Mechanics 
}

\author{Tsunehiro Kobayashi\footnote{E-mail: 
kobayash@a.tsukuba-tech.ac.jp} \\
{\footnotesize\it Department of General Education 
for the Hearing Impaired,}
{\footnotesize\it Tsukuba College of Technology}\\
{\footnotesize\it Ibaraki 305-0005, Japan}}

\date{}

\maketitle

\begin{abstract}

Vortex lattices are constructed  
in terms of linear combinations of solutions 
for Scr\"{o}dinger equation with a constant 
potential. 
The vortex lattices are mapped on the spaces with two-dimensional 
rotationally symmetric potentials by using conformal mappings 
and the differences of the mapped vortex-patterns are examined. 
The existence of vortex dipole and quadrupole is also pointed out. 

\end{abstract}

\thispagestyle{empty}

\setcounter{page}{0}

\pagebreak

Vortices play an interesting roles in various aspects of 
present-day physics such as vortex matters (vortex lattices) 
in condensed matters [1,2], 
quantum Hall effects
 [3-7], 
various vortex patterns of 
non-neutral plasma [8-11] 
and Bose-Einstein gases [12-16]. 
The vortices are well known objects in hydrodynamics and 
has been investigated in many aspects [17-21]. 
In quantum mechanics hydrodynamical approach was vigorously 
investigated in the early stage of the development 
of quantum mechanics [22-29]. 
The fundamental properties of vortices in quantum mechanics 
were extensively examined by Hirschfelder and 
others [30-34] 
and the motions of vortex lines were also studied [35,36]. 
Recently an analysis of vortices has been carried out on the 
basis of non-linear Schr\"{o}dinger equations [37,38]. 
The vortex dynamics will hold an important position in 
quantum mechanical problems. 
In vortex dynamics, however, the construction of various vortex patterns 
from fundamental dynamics is still not very clear. 
Recently Kobayashi and Shimbori has proposed a way to investigate 
vortex patterns from the special solutions of Schr\"{o}dinger 
equations [39]. 
By using the conformal mappings the article has shown that 
the special solutions with zero energy eigenvalue in the 
two dimensional parabolic potential barriers (2D PPB) 
expressed by 
$V=-m\gamma^2(x^2+y^2)/2$ [40] 
can be extended 
to all rotationally symmetric potentials 
and the solutions are infinitely degenerate as same as 
those in the 2D PPB [39]. 
It has also been shown that the infinite degeneracy can be 
an origin of variety of vortex patterns which appear at nodal points of 
wave functions and 
this idea can be applied in three dimensions. 
In the article a few examples of simple vortex patterns are presented. 
In this letter some patterns of vortex lattices that are constructed from 
simple linear combinations of the infinitely degenerate states 
will be studied. 

Let us start fom the arguments in two dimensions [39]. 
In two dimensions 
the eigenvalue problems with the energy eigenvalue ${\cal E}$ 
are explicitly written as
\begin{equation}
 [-{\hbar^2 \over 2m}\vtri +V_a(\rho)]\ \psi(x,y) 
  = {\cal E}\ \psi(x,y),
  \label{1}
\end{equation}
where 
$
\vtri=\partial^2/ \partial x^2+\partial^2 / \partial y^2,
$  
 the rotationally symmetric potentials are generally given by  
$$
V_a(\rho)=-a^2 g_a\rho^{2(a-1)},
$$ 
with $\rho=\sqrt{x^2+y^2}$, $a \in \real$ ($a\not=0$), and 
$m$ and $g_a$ are, respectively, the mass of the particle and the coupling 
constant. 
Note here that the eigenvalues ${\cal E}$ should generally 
be taken as complex numbers 
that are allowed only in conjugate spaces of Gel'fand triplets [41]. 
Note also that $V_a$ represents repulsive potentials for $(g_a>0,\ a>1)$ and 
$(g_a<0,\ a<1)$ and attractive potentials for 
$(g_a>0,\ a<1)$ and $(g_a<0,\ a>1)$. 

Let us consider the conformal mappings 
\begin{equation}
\zeta_a=z^a,\ \ \ \ \ \  {\rm with}\ z=x+iy.
\label{2}
\end{equation} 
We use the notations $u_a$ and $v_a$ defined by 
$
\zeta_a=u_a+iv_a
$ 
that are written as  
\begin{equation} 
u_a=\rho^a\cos a\varphi,\ \ v_a=\rho^a\sin a\varphi,
\label{3}
\end{equation} 
where $\varphi=\arctan (y/x)$. 
In the $(u_a,v_a)$ plane the equations (1) are written down as 
\begin{equation}
a^2\rho_a^{2(a-1) / a} [-{\hbar^2 \over 2m}\vtri_a-g_a]\ \psi(u_a,v_a)=
     {\cal E}\ \psi(u_a,v_a), 
 \label{4}
\end{equation}
where 
$ 
\vtri_a=\partial^2/ \partial u_a^2+\partial^2/ \partial v_a^2.
$ 
We can rewrite the equations as
\begin{equation}
[-{\hbar^2 \over 2m}\vtri_a-g_a]\ \ \psi(u_a,v_a)=
    a^{-2}{\cal E}\ \rho_a^{2(1-a) / a} \psi(u_a,v_a). 
 \label{5}
\end{equation} 
It is surprising that the equations become same for all $a\not=0$ when 
the energy eigenvalue ${\cal E}$ have zero value. 
That is to say, for ${\cal E}=0$ the equations have the same form as that for 
the free particle  with the constant potentials $g_a$ as 
\begin{equation}
[-{\hbar^2 \over 2m}\vtri_a-g_a]\ \ \psi(u_a,v_a)=0. 
 \label{6}
\end{equation} 
It should be noticed that in the case of $a=1$ where the original potential 
is a constant $g_1$ the energy does not need to be zero but can take arbitrary 
real numbers, because the right-hand side of (5) has no $\rho$ dependence.  
In the $a=1$ case, therefore, we should take $g_1+{\cal E}$ instead of 
$g_a$. 
Here let us briefly comment on the conformal mappings $\zeta_a=z^a$. 
We see that the transformation maps the part of the $(x,y)$ plane 
described by $0\leq \rho<\infty, \ 0\leq \varphi<\pi/a$ 
on the upper half-plane of the $(u_a,v_a)$ plane for $a>1$ 
and the lower half-plane for $a<-1$. 
Note here that the maps on the part of the $(u_a,v_a)$ plane with 
the angle $\varphi_a=\varphi+\alpha$ can be carried out 
by using the conformal mappings 
\begin{equation}
\zeta_a(\alpha)=z^a e^{i\alpha}.
\label{7}
\end{equation}
In the maps the variables are given by 
\begin{equation}
u_a(\alpha)=u_a\cos\alpha - v_a\sin\alpha, \ \ \ 
v_a(\alpha)=v_a\cos\alpha + u_a\sin\alpha. 
\label{8}
\end{equation} 
The relations $u_a(0)=u_a$ and $v_a(0)=v_a$ are obvious. 

It is trivial that the equations for all $a$ have the 
particular solutions 
\begin{align}
\psi_0^\pm(u_a)&=N_a e^{\pm ik_au_a}    \label{9}\\ 
\psi_0^\pm(v_a)&=N_a e^{\pm ik_av_a}, \ \ {\rm for}\ g_a>0 \label{10}
\end{align} 
and 
\begin{align}
\phi_0^\pm(u_a)&=M_a e^{\pm k_au_a}  \label{11} \\
\phi_0^\pm(v_a)&=M_a e^{\pm k_av_a}, \ \ {\rm for}\ g_a<0 \label{12}
\end{align} 
where $k_a=\sqrt{2m|g_a|}/\hbar$, and $N_a$ and $M_a$ are 
in general complex numbers. 
General solutions should be written by the linear combinations 
of (9) and (10) for $g_a>0$ and 
those of (11) and (12) for $g_a<0$. 
Examples for the 2D PPB with $g_a>0$ are presented in ref. [40]. 
In the following investigations we shall concentrate our attention on 
 the solutions of (9) and (10) 
that are expressed in terms of plane waves. 
Hereafter we use the notations 
$u_a(\alpha)$ and 
$v_a(\alpha)$ 
with 
$-\pi\leq \alpha <\pi$, which are used in the conformal mappings of (7). 
We see that 
$$
e^{\pm ik_au_a(\alpha )}\ \ {\rm and} \ \ e^{\pm ik_av_a(\alpha )}
$$ 
are the solutions of (6). 
(For details, see ref. [39].) 

We have to comment on the degeneracy of the solutions. 
The origin of the infinite degeneracy can easily understand 
in the case of the 2D PPB [40]. 
It is known that 
energy eigenvalues of 1D PPB are given by pure imaginary values 
$\mp i(n+1/2)\hbar \gamma$ with 
$n=0,1,2,\cdots$ [42-47]. 
From this reult we see that the energy eigenvalues of the 2D PPB, 
which are composed of the sum of the 1D-PPB eigenvalues with 
the oposite signs such as 
 $i(n_x-n_y)\hbar\gamma$ with $n_x$ and $n_y=0,1,2,\cdots$, 
 include the zero energy for $n_x=n_y$. 
It is apparent that all the states with the energy eigenvalues
$i(n_x-n_y)\hbar\gamma$ are infinitely degenerate [40]. 
This means that all the rotationally symmetric potentials have 
the same degeneracy for the zero energy states.  
By putting the wave function $f^\pm (u_a;v_a)\psi_0^\pm(u_a)$ 
into (6) 
where $f^\pm (u_a;v_a)$ is a polynomial function of $u_a$ and $v_a$, 
we obtain the equation 
\begin{equation}
[\vtri_a \pm2ik_a{\partial \over \partial u_a}]f^\pm (u_a;v_a)=0.
\label{13}
\end{equation} 
A few examples of the functions $f$ 
are given by [40] 
\begin{align}
f_0^\pm (u_a;v_a)&=1, \nonumber \\
f_1^\pm (u_a;v_a)&=4k_a v_a, \nonumber \\
f_2^\pm (u_a;v_a)&=4(4k_a^2 v_a^2+1\pm 4 i k_a u_a).
\label{14}
\end{align} 
We can obtain the general forms of the polynomials in the 2D PPB, 
which are generally written by the multiple of 
the polynomials of degree $n$, $H_n^\pm(\sqrt{2k_2} x)$, 
such that 
\begin{equation}
f_{n}^{\pm}(u_2;v_2)=H_{n}^\pm(\sqrt{2k_2} x)
\cdot H_{n}^\mp(\sqrt{2k_2} y),
\label{15} 
\end{equation} 
where $x$ and $y$ in the right-hand side should be 
considered as the functions of $u_2$ and $v_2$ [40]. 
Since the form of the equations (6)  
is the same for all $a$, 
the solutions can be written by the same polynomial functions 
that are given in (15) for the PPB. 
That is to say, 
we can obtain the polynomials for arbitrary $a$ 
by replacing $u_2$ and $v_2$ with $u_a$ and $v_a$ 
in (15). 
Note that the polynomials $H_{n}^\pm(\xi)$ with 
$\xi=\sqrt{m\gamma/\hbar} x$ 
are defined 
by the solutions for the eigenstates with 
${\cal E}=\mp i(n+1/2)\hbar \gamma$ in 1D 
PPB of the type $V(x)=-m\gamma^2x^2/2$ and they are 
written in terms of the Hermite polynomials $H_{n}(\xi)$ as 
\begin{equation}
H_{n}^\pm(\xi)=e^{\pm i n\pi/4} H_{n}(e^{\mp i\pi/4}\xi). 
\label{16}
\end{equation} 
(For details, see ref. [45,46].) 
The states expressed by these wave functions belong to the conjugate 
spaces of Gel'fand triplets of which nuclear space is given by 
Schwarz space. 
Actually we easily see that the wave functions cannot be normalized 
in terms of Dirac's delta functions except the lowest polynomial 
solutions [39]. 

The extension to three dimensions can easily be carried out in 
the cases with potentials that are separable into the $(x,y)$ plane 
and the $z$ direction such that 
$$
V(x,y,z)=V_a(\rho)+V(z).
$$ 
When the energy eigenvalues of the $z$ direction is given by $E_z$, 
we obtain the same equation as (6) for ${\cal E}-E_z=0$. 
(For the $a=1$ case $g_1+{\cal E}-E_z>0$ should be taken.) 
If we take the free motion with the momentum $p_z$ for the $z$ direction, 
$E_z=p_z^2/2m$ should be taken. 
It is important that the total energy ${\cal E}$ is in general not equal to 
zero in the three dimensions. 
Note that wave functions for the separable potentials are written by 
the product such as 
$\psi(x,y,z)=\psi(x,y)\psi(z)$. 
Hereafter we shall not explicitly write $\psi(z)$ in the wave functions. 

Let us study vortices that appear in the linear combinations composed of 
the solutions with the polynomials of (15). 
Before going into the details we briefly describe vortices 
in quantum mechanical hydrodynamics. 
The probability density $\rho(t,x,y)$ and 
the probability current $\vj(t,x,y)$ of a wave function $\psi(t,x,y)$ 
in non-relativistic quantum mechanics 
are defined 
by 
\begin{align}
  \rho(t,x,y)&\equiv\left| \psi(t,x,y)\right|^2, \label{6.1.1}\\
  \vj(t,x,y)&\equiv{\rm Re}\left[\psi(t,x,y)^*
  \left(-i\hslash\nabla\right)\psi(t,x,y)\right]/m.
  \label{17}
\end{align}
They satisfy the equation of continuity 
 $ \partial\rho/\partial t+\nabla\cdot\vj=0.$ 
Following the analogue of the hydrodynamical 
approach [17-21], 
the fluid 
can be represented by 
the density $\rho$ and the fluid velocity $\vv$. 
They satisfy Euler's equation of continuity 
\begin{equation}
  \frac{\partial\rho}{\partial t}+\nabla\cdot(\rho\vv)=0. 
   \label{18}
\end{equation} 
Comparing this equation with the continuity equation, 
we are thus led to the following 
definition for the quantum velocity of the state $\psi(t,x,y)$, 
\begin{equation}
  \vv\equiv\frac{\vj(t,x,y)}{\left| \psi(t,x,y)\right|^2}, 
   \label{19}
\end{equation}
in which $\vj(t,x,y)$ is given by (17). 
Notice that $\rho$ and $\vj$ in the present cases 
do not depend on time $t$. 
Now it is obvious that vortices appear at the zero points of 
the density, that is, the nodal points of the wave function, 
where the current $\vj$ does not vanish [39]. 
We should here remember that the solutions of (6) 
degenerate infinitely. 
This fact indicates that we can construct wave functions having 
the nodal points at arbitrary positions in terms of 
linear combinations of the infinitely degenerate 
solutions [39,40]. 

The strength of vortex is characterized 
by the circulation $\Gammav$ 
that is represented by the integral round a closed contour $C$
encircling the vortex such that 
\begin{equation}
\Gammav=\oint_C \vv \cdot d\vs
\label{20}
\end{equation} 
and it is quantized as 
\begin{equation}
\Gammav=2\pi l\hbar/m, 
\label{21}
\end{equation} 
where the circulation number $l$ is 
an integer [31,33,36]. 
It should be stressd that 
we can perform the investigation of vortices 
for all the cases with $a\not=0$ in the $(u_a,v_a)$ plane, 
because fundamental properties of vortices such as 
the numbers of vortices in the original plane and the mapped plane and 
the strengths of vortices do not change by the conformal mappings. 
Vortex patterns in the $(x,y)$ plane can be obtained by the inverse 
transformations of the conformal mappings. 
Let us here show that vortex lines and vortex lattices can be 
constructed from simple linear combinations of the low lying  
polynomial solutions. 
And also the mapped patterns of those lines and lattices 
are investigated by the conformal mappings for the $a=2$ 
(PPB case; $V_a\propto \rho^2$) 
and for the $a=1/2$ (Coulomb type; $V_a\propto \rho^{-1}$). 
In the following discussions the suffices $a$ of $u_a$, $v_a$ and $k_a$ 
are omitted. 

\hfil\break
(I) Vortex lines

Let us consider the linear combination of two degree 1 solutions such that 
\begin{equation}
\Psi(u,v)=ve^{iku}-ue^{-ikv}, 
\label{22}
\end{equation}
where the complex constant corresponding to the overall factor of 
the wave function is ignored, 
because the wave function belongs to the conjugate space of Gel'fand triplet 
and it is not normalizable. 
This means that the wave function represents a stationary flow [40]. 
The nodal points of the probability density 
$$
|\Psi(u,v)|^2=u^2+v^2-2uv \cos k(u+v)
$$ 
appear at points satisfying the conditions 
\begin{equation}
u=\pm v, \ \ \ \ \cos k(u+v)=\pm 1. 
\label{23}
\end{equation} 
We have the nodal points at 
\begin{equation}
u=v=n\pi /k, \ \ \ {\rm for }\ n={\rm integers}. 
\label{24}
\end{equation} 
In the $(u,v)$ plane the positions of vortices can be on a line 
of $u=v$. 
After some elementary but tedious calculations we see that 
the circulation numbers of vortex strengths are given by $l=-1$ for 
$n=$positive integers and $l=1$ for $n=$negative ones. 
Note that the origin at $u=v=0$ has no vortex. 
We can directly 
see the result by showing the fact that 
the strength of vortex $\varGamma$ 
becomes zero for the closed circle around the origin. 
We can interpret this result as follows; at the origin there exist 
a pair of vortices having the opposite circulation numbers, that is, 
they, respectively, belong to the vortex line with $l=-1$ and 
that with $l=1$. 
We may say that it is a vortex dipole. 

For the case of $a=1$ (constant potential) we can take as $u=x$ and 
$v=y$. 

In the case of $a=2$ (PPB) we have 
$u=x^2-y^2$ and $v=2xy$.  
The relations for the nodal points 
are written down as 
\begin{align}
y&={1 \over \sqrt{2}+1}x,\ \ y=\pm {1 \over 2}\sqrt{n\pi \over (\sqrt{2}+1)k}, \ \ \ 
{\rm for}\ n={\rm positive\ integers} \nonumber \\
y&=-{1 \over \sqrt{2}-1}x,\ \ y=\pm {1 \over 2}\sqrt{|n|\pi \over (\sqrt{2}-1)k}, \ \ \ 
{\rm for}\ n={\rm negative\ integers}. \nonumber \\
\label{25}
\end{align} 
We see that a vortex quadrupole composed of two vortex dipole 
appears at the origin. 

In the case of $a=1/2$ (Coulomb type), 
by using the relations $u^2-v^2=x$ and $2uv=y$, we obtain 
the conditions for the nodal points as follows; 
\begin{equation}
x=0,\ \ \ y=2{n^2\pi^2 \over k^2}, \ \ \ 
{\rm for}\ n={\rm non\ zero\ integers}. 
\label{26}
\end{equation}  
Note that the origin is a singular point, where the source of 
the potential exists. 

Figures for $a=1,2$ and $1/2$ are presented 
in figs.1, 2 and 3, which, 
respectively, represent the vortex pattern for the constant potential, 
that for the PPB, and that for the Coulomb type one. 
Note here that the differences of the potentials clearly appear 
not only in the vortex patterns but the properties of 
the singularities at the origin as well. 
In hydrodynamics it is known that 
vortex lines for the constant potential 
($a=1$) are stationary but generally unstable 
for perturbations [17-21]. 
Note that the parallel vortex lines are constructed from the linear 
combinations of the lowest and the degree 1 polynomials [39]. 

\hfil\break
(II) Vortex lattices

Let us consider the linear combination of a stationary wave and a plane 
wave such that 
\begin{equation}
\Psi(u,v)=\cos ku - e^{-ikv}.
\label{27}
\end{equation} 
The nodal points of the probability density 
$$
|\Psi(u,v)|^2=1+\cos^2 ku - 2\cos ku \cos kv
$$ 
appear at positions satisfying 
\begin{equation}
u=m\pi/k, \ \ \ \ v=n\pi/k,
\label{28}
\end{equation} 
where both of $m$ and $n$ must be even or odd, that is, 
$(-1)^m=(-1)^n$ must be fulfilled. 
These conditions produce a vortex lattice presented 
in fig.4, which was suggested 
in ref.[48]. 

In the cases of the PPB ($a=2$) and the Coulomb type ($a=1/2$) 
vortices appear at the cross points 
of the following two functions; 
\begin{align}
x^2-y^2&=m\pi/k, \ \ \ \ xy=n\pi/2k,\ \ \ {\rm for\ the\ PPB}, \nonumber \\
x^2+y^2&=(m^2+n^2)^2\pi^4/k^4, \ \ \ \ y=2mn\pi^2/k^2,\ \ \ 
{\rm for\ the\ Coulomb\ type}.
\label{29}
\end{align} 
In these case we obtain the circulation number $l=-1$ 
for the all vortices. 
Figures for $a=2$ and $1/2$ are given 
in figs.5 and 6, respectively. 

In these arguments we see the following points: 
\hfil\break
(1) The construction of vortex lattices in experiments seems to be 
not very difficult. In fact the vortex lattice of (II) can be produced 
from a stationary wave and a plane wave perpendicular to the 
stationary wave. 
\hfil\break
(2) The differences of potentials can be clearly seen from the vortex 
patterns. 
Especially the distances between two neighbouring vortices 
are a good object to identify 
the type of the potentials. 
That is to say, the vortices appear in an equal distance $\pi/k$ 
in the case of the constant potential ($a=1$), 
whereas the distances become smaller in the regions far from the 
origin for $a>1$ and larger for $a<1$ 
in comparison with those near the origin. 
\hfil\break
(3) The property of the singularity at the origin is also a good 
object to identify the potentials. 
\hfil\break
Though it is at this moment difficult to categorize the 
present experimental vortex patterns [8-16], 
we shall be able to understand fundamental dynamics of vortex phenomena 
from vortex patterns. 

It is also noticed that the present results can be applicable not only 
to quantum phenomena but also those in classical fluids by changing 
the parameters $m$, $\hbar$ and $g_a$ in the original equation. 

Up to now we have not discussed 
on the stability of the vortex lattices. 
In order to investigate the time development of the patterns we have to 
take account of the fact that the solutions used here belong to 
the conjugate spaces of Gel'fand triplets. 
In the spaces the eigenstates generally have complex energy eigenvalues 
such as resonances and 
the eigenvalues are expressed by pairs of complex conjugates such that 
${\cal E}=\epsilon \pm i\gamma$, 
where $\epsilon,\gamma \in \real$ [41]. 
It is known that the + and - signs of the imaginary part in the eigenvalues, 
respectively, represent resonance-formation and resonance-decay processes.  
We see that this pairing property is 
the origin of the infinite degeneracy of the solutions and 
the infinite degeneracy stems 
from the balance between the resonance-formation 
and resonance-decay processes. 
This fact seems to indicate 
that many vortex systems are possibly unstable for perturbations. 
Actually the existence of vortex lattices has already been pointed 
out and it has also been noticed that  
those systems will decay from their edges, 
where the balance between the formation and decay processes is 
broken [48]. 
In such processes we have to discuss on dynamics in many vortex systems 
and statistical mechanics should be extended into 
the conjugate spaces of Gel'fand triplets, 
where the freedom of the imaginary energy eigenvalues must be 
introduced [48-50]. 
In the theory the infinite degeneracy plays another important role, 
that is, it becomes the origin of new entropy. 
The new entropy brings essentially new processes in thermal 
non-equilibrium [49,50]. 
At present we still have many fundamental questions in the investigation 
of dynamics in the conjugate spaces of Gel'fand triplets. 
The study of vortex dynamics will brings us a new prospect and open 
a new dynamics for the systems essentially 
described by the states in Gel'fand triplets.

\hfil\break 
  [1] 
           Blatter~G, et al., 
           \emph{Rev. Mod. Phys.} {\bf 66} (1994) 1125.  
\hfil\break 
  [2] 
          Crabtree~G W and Nelson~D R, 
          \emph{Phys. Today} {\bf 50} No.4 (1997) 38. 
\hfil\break 
  [3] 
          Prange R E and Girvin M, 
          \emph{The Quantum Hall Effect} 2nd ed 
          (Springer) 1990.
 \hfil\break          
  [4] 
          Wilczek F, 
          \emph{Fractional Statistics and Anyon Superconductivity}
          (World Scientific) 1990.
 \hfil\break          
  [5] 
          Chakaraborty T and Pietil\"{a}inen P, 
          \emph{The Quantum Hall Effects: Fractional and 
          Integral} 2nd and updated ed 
          (Springer) 1995.
 \hfil\break          
  [6] 
          Das Sarma S and Pinczuk A, 
          \emph{Perspectives in Quantum Hall Effects}
          (Wiley) 1997.
\hfil\break           
  [7] 
          Khare A, 
          \emph{Fractional Statistics and Quantum Theory}
          (World Scientific) 1997.
\hfil\break 
  [8] 
          Fine~K S, et al., 
          \emph{Phys. Rev. Lett.} {\bf 75} (1995) 3277.
\hfil\break 
  [9] 
          Kiwamoto~Y, et al., 
          \emph{J. Phys. Soc. Jpn. (Lett.)} {\bf 68} (1999) 3766.
\hfil\break 
 [10] 
          Kiwamoto~Y, et al., 
          \emph{Phys. Rev. Lett.} {\bf 85} (2000) 3173.
\hfil\break 
  [11] 
          Ito~K, et al.,  2001 
          \emph{Jpn. J. Appl. Phys.} (2001) (to appear).
\hfil\break 
  [12] 
          Matthews~R M, et al., 
          \emph{Phys. Rev. Lett.} {\bf 83} (1999) 2498.
\hfil\break 
  [13] 
          Raman~C, et al., 
          \emph{Phys. Rev. Lett.} {\bf 83} (1999) 2502.
\hfil\break 
  [14] 
          Madison~K W, et al., 
          \emph{Phys. Rev. Lett.} {\bf 84} (2000) 806.
\hfil\break 
  [15] 
          Marago~O M, et al., 
          \emph{Phys. Rev. Lett.} {\bf 84} (2000) 2056. 
\hfil\break 
  [16] 
          Fitzgerald~R, et al., 
          \emph{Phys. Today} {\bf 55} No.8 (2000) 19.
\hfil\break 
  [17] 
          Lamb~H, 
          \emph{Hydrodynamics} 6th ed 
          (Cambridge: Cambridge Univ. Press) 1932. 
  \hfil\break 
  [18] 
          Landau~L~D and Lifshitz~E~M, 
          \emph{Fluid Mechanics} 2nd ed 
          (Oxford: Pergamon) 1987.
 \hfil\break  
  [19] 
          Batchelor~G~K, 
          \emph{An Introduction to Fluid Dynamics} 
          (Cambridge: Cambridge Univ. Press) 1967.
  \hfil\break          
  [20] 
          Tatsumi~T, 
          \emph{Hydrodynamics} (in Japanese) 
          (Tokyo: Baihuukann) 1982.
   \hfil\break 
  [21] 
          Saffman~P~G, 
          \emph{Vortex Dynamics} 
          (Cambridge: Cambridge Univ. Press) 1992.
   [22] 
          Madelung~E, 
          \emph{Z. Phys.} {\bf 40} (1926) 322.
  \hfil\break 
  [23] 
          Kennard~E~H, 
          \emph{Phys. Rev.} {\bf 31} (1928) 876. 
  \hfil\break 
  [24] 
          de~Broglie~L, 
          \emph{Introduction \`{a} l'\'{e}tude de la 
          M\'{e}canique ondulatoire} 
          (Paris: Hermann) 1930.
  \hfil\break 
  [25] 
          Dirac~P~A~M 
           \emph{Proc. Roy. Soc. (London)} {\bf A209} (1951) 291; 
           \emph{Proc. Roy. Soc. (London)} {\bf A212} (1952) 330; 
           \emph{Proc. Roy. Soc. (London)} {\bf A223} (1954) 438. 
  \hfil\break 
  [26] 
          Bohm~D 
           \emph{Phys. Rev.} {\bf 85} (1952) 166 and 180; 
           \emph{Phys. Rev.} {\bf 89} (1953) 458. 
  \hfil\break 
  [27] 
          Takabayasi~T 
           \emph{Prog. Theor. Phys.} {\bf 8} (1952) 143; 
           \emph{Prog. Theor. Phys.} {\bf 9} (1953) 187. 
  \hfil\break 
  [28] 
          Sch\"{o}nberg~M, 
          \emph{Nuovo Cim.} {\bf 12} (1954) 103. 
  \hfil\break 
  [29] 
          Bohm~D and Vigier~J~P 
           \emph{Phys. Rev.} {\bf 96} (1954) 208; 
           \emph{Phys. Rev.} {\bf 109} (1958) 1882. 
  \hfil\break 
   [30] 
          Hirschfelder~J~O, Christoph~A~C and Palke~W~E, 
          \emph{J. Chem. Phys.} {\bf 61} (1974) 5435. 
  \hfil\break 
  [31] 
          Hirschfelder~J~O, Goebel~C~J and Bruch~L~W, 
          \emph{J. Chem. Phys.} {\bf 61} (1974) 5456. 
  \hfil\break 
  [32] 
          Hirschfelder~J~O and Tang~K~T, 
          \emph{J. Chem. Phys.} {\bf 64} (1976) 760; {\bf 65} 
          (1976) 470. 
  \hfil\break 
  [33] 
          Hirschfelder~J~O, 
          \emph{J. Chem. Phys.} {\bf 67} (1977) 5477. 
  \hfil\break  
  [34] 
          Ghosh~S~K and Deb~B~M, 
          \emph{Phys. Reports} {\bf 92} (1982) 1. 
 \hfil\break  
  [35] 
          Schecter~D~A and Dubin~H~E, 
          \emph{Phys. Rev. Lett.} {\bf 83} (1999) 2191. 
 \hfil\break          
 [36] 
          Bialynicki-Birula~I, Bialynicka-Birula~Z 
          and \'{S}liwa~C, 
          \emph{Phys. Rev.} {\bf A61} (2000) 032110. 
 \hfil\break 
 [37] 
          Dmitriyev~V~P, 
          \emph{Z. Naturforsch.} {\bf 48} (1993) 935. 
\hfil\break  
 [38] 
          Dmitriyev~V~P, 
          Mechanical analogy for the wave-particle: helix on 
          a vortex filament 
          \emph{Preprint} quant-ph/0012008 (2001). 
\hfil\break 
  [39] 
          Kobayashi~T and Shimbori~T, 
          Zero Energy Solutions and Vortices in Schr\"{o}dinger 
          Equations
          \emph{Preprint} cond-mat/0103209 (2001). 
  \hfil\break 
  [40] 
          Shimbori~T and Kobayashi~T, 
          \emph{J. Phys.} {\bf A33} (2000) 7637.
\hfil\break 
 [41] 
	  Bohm~A and Gadella~M, 
	  \emph{Dirac Kets, Gamow Vectors and Gel'fand Triplets} 
	  (Lecture Notes in Physics, Vol. 348, Springer, 1989). 
 \hfil\break 
  [42] 
          Barton~G, 
          \emph{Ann. Phys.} {\bf 166} (1986) 322.
  \hfil\break 
  [43] 
          Briet~P, Combes~J~M and Duclos~P, 
          \emph{Comm. Partial Differential Equations} {\bf 12} 
          (1987) 201. 
  \hfil\break 
  [44] 
          Balazs~N~L and Voros~A, 
          \emph{Ann. Phys.} {\bf 199} (1990) 123. 
  \hfil\break 
  [45] 
          Castagnino~M, Diener~R, Lara~L and Puccini~G, 
          \emph{Int. J. Theor. Phys.} {\bf 36} (1997) 2349. 
  \hfil\break 
  [46] 
  Shimbori~T and Kobayashi~T, 
  \emph{Nuovo Cim.} {\bf 115B} (2000) 325. 
  \hfil\break 
  [47] 
  Shimbori~T, 
          \emph{Phys. Lett.} {\bf A273} (2000) 37. 
  \hfil\break 
  [48] 
          Kobayashi~T and Shimbori~T, 
          Statistical mechanics 
          for states with complex eigenvalues 
          and quasi-stable semiclassical systems 
          \emph{Preprint} cond-mat/0005237 (2000).
\hfil\break 
  [49] 
          Kobayashi~T and Shimbori~T, 
          \emph{Phys. Lett.} {\bf A280} (2001) 23.
 \hfil\break         
 [50] 
          Kobayashi~T and Shimbori~T, 
          \emph{Phys. Rev. {\bf E}} to appear on 1, May (2001) 
          (cond-mat/0009142). 


\pagebreak

\begin{figure}[pb]
   \begin{center}
    \begin{picture}(200,200)
     \thicklines
     \put(0,100){\vector(1,0){200}}
     \put(100,10){\vector(0,1){190}}
     \put(205,98){$x$}
     \put(98,205){$y$}
     
     \put(10,10){\line(1,1){190}}
     
          \put(97.7,97.7){$\circ$}

     \put(30,25){$-3$}
     \put(55,50){$-3$}
     \put(80,75){$-1$}
     \put(132,123){$1$}
     \put(157,148){$2$}
     \put(182,173){$3$}
     
     \put(25,25){$\bullet$}
     \put(50,50){$\bullet$}
     \put(75,75){$\bullet$}
     \put(125,125){$\bullet$}
     \put(150,150){$\bullet$}
     \put(175,175){$\bullet$}
    \end{picture}
   \end{center}
   \caption[]{Positions of vortices for $n=\pm 1,\ \pm 2,\ \pm 3$ 
   in the constant potential ($a=1$), 
   which are denoted by $\bullet$, 
   and $\circ$ stands for the vortex dipole at the origin.} 
   \label{fig:1.1}
  \end{figure}
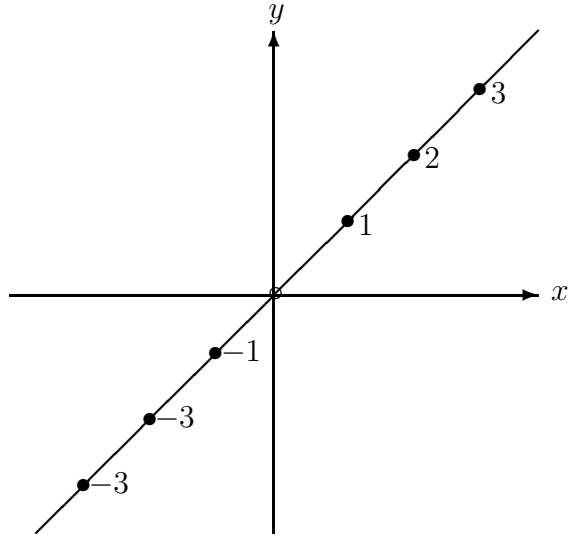 

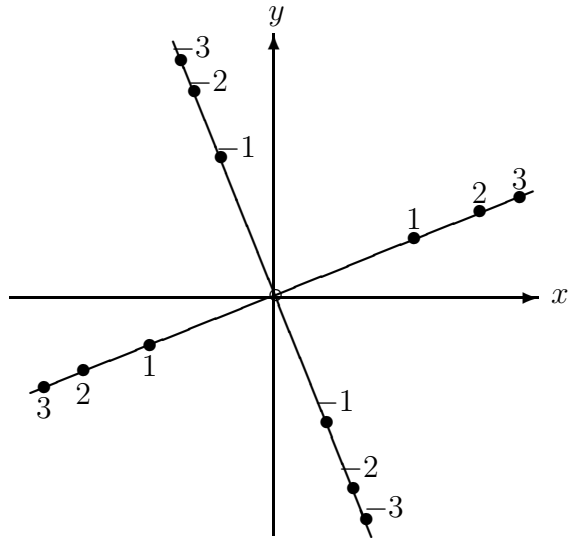
\begin{figure}
   \begin{center}
    \begin{picture}(200,200)
     \thicklines
     \put(0,100){\vector(1,0){200}}
     \put(100,10){\vector(0,1){190}}
     \put(205,98){$x$}
     \put(98,205){$y$}
     
     \put(8,63.93){\line(5,2){190}}
     \put(61.97,197){\line(2,-5){75}}
     
          \put(97.8,97.8){$\circ$}

     \put(10,54.53){$3$}
     \put(25,59.53){$2$}
     \put(50,70.29){$1$}
     \put(150,126.01){$1$}
     \put(175,135.60){$2$}
     \put(190,141){$3$}
     
     \put(134,17){$-3$}
     \put(124.507,32){$-2$}
     \put(115.271,57){$-1$}
     \put(78.09,153){$-1$}
     \put(67.53,178){$-2$}
     \put(61,191){$-3$}

     \put(10,63.1){$\bullet$}
     \put(25,69.43){$\bullet$}
     \put(50,79.09){$\bullet$}
     \put(150,119.51){$\bullet$}
     \put(175,129.67){$\bullet$}
     \put(190,134.82){$\bullet$}
     \put(62,187){$\bullet$}
     \put(127.07,25){$\bullet$}
     \put(117.01,50){$\bullet$}
     \put(77.05,150){$\bullet$}
     \put(67.08,175){$\bullet$}
     \put(132.1,13){$\bullet$}
    
    \end{picture}
   \end{center}
   \caption[]{Positions of vortices for $n=\pm 1,\ \pm 2,\ \pm 3$ 
   in the PPB ($a=2$), 
   which are denoted by $\bullet$, 
   and $\circ$ stands for the vortex quadrupole at the origin.} 
   \label{fig:1.2}
   \end{figure}

   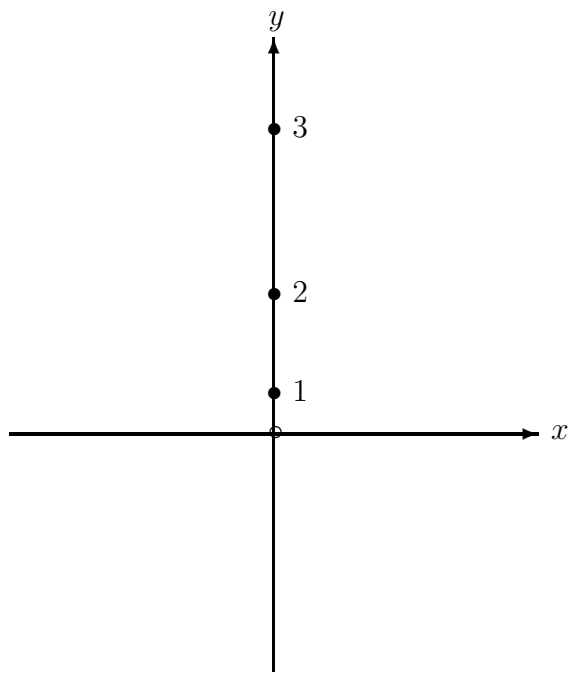
\begin{figure}
   \begin{center}
    \begin{picture}(200,250)
     \thicklines
     \put(0,100){\vector(1,0){200}}
     \put(100,10){\vector(0,1){240}}
     \put(205,98){$x$}
     \put(98,255){$y$}
     
     \put(107,112.5){$1$}
     \put(107,150){$2$}
     \put(107,212.5){$3$}
     
     \put(97.7,97.8){$\circ$}
     \put(97.3,112.5){$\bullet$}
     \put(97.3,150){$\bullet$}
     \put(97.3,212.5){$\bullet$}
    \end{picture}
   \end{center}
   \caption[]{Positions of vortices for $n=1,\ 2,\ 3$ 
   in the Coulomb type potential ($a=1/2$), 
   which are denoted by $\bullet$, 
   and $\circ$ stands for the source at the origin.} 
   \label{fig:1.3}
  \end{figure}

 \begin{figure}[pb]
   \begin{center}
    \begin{picture}(300,300)
     \thicklines
     \put(0,152){\vector(1,0){300}}
     \put(152,10){\vector(0,1){290}}
     \put(305,150){$x$}
     \put(150,305){$y$}
     \put(137,0){$m=0$}
     \put(0,155){$n=0$}
     
     \put(0,32.5){\line(1,0){290}}
     \put(0,62.5){\line(1,0){290}}
     \put(0,92.5){\line(1,0){290}}
     \put(0,122.5){\line(1,0){290}}
     \put(0,182.5){\line(1,0){290}}
     \put(0,212.5){\line(1,0){290}}
     \put(0,242.5){\line(1,0){290}}
     \put(0,272.5){\line(1,0){290}}
    
     \put(32.5,15){\line(0,1){270}}
     \put(62.5,15){\line(0,1){270}}
     \put(92.5,15){\line(0,1){270}}
     \put(122.5,15){\line(0,1){270}}
     \put(182.5,15){\line(0,1){270}}
     \put(212.5,15){\line(0,1){270}}
     \put(242.5,15){\line(0,1){270}}
     \put(272.5,15){\line(0,1){270}}
     
     
     \put(0,290){$m=-4$}
     \put(56,290){$-3$}
     \put(86,290){$-2$}
     \put(116,290){$-1$}
     \put(176,290){$1$}
     \put(206,290){$2$}
     \put(236,290){$3$}
     \put(266,290){$4$}
     
     \put(295,30){$-4$}
     \put(295,60){$-3$}
     \put(295,80){$-2$}
     \put(295,120){$-1$}
     \put(295,180){$1$}
     \put(295,210){$2$}
     \put(295,240){$3$}
     \put(285,275){$n=4$}

     \put(30,30){$\bullet$}
     \put(30,90){$\bullet$}
     \put(30,150){$\bullet$}
     \put(30,210){$\bullet$}
     \put(30,270){$\bullet$}
     \put(60,60){$\bullet$}
     \put(60,120){$\bullet$}
     \put(60,180){$\bullet$}
     \put(60,240){$\bullet$}
     \put(90,30){$\bullet$}
     \put(90,90){$\bullet$}
     \put(90,150){$\bullet$}
     \put(90,210){$\bullet$}
     \put(90,270){$\bullet$}
     \put(120,60){$\bullet$}
     \put(120,120){$\bullet$}
     \put(120,180){$\bullet$}
     \put(120,240){$\bullet$}
     \put(150,30){$\bullet$}
     \put(150,90){$\bullet$}
     \put(150,150){$\bullet$}
     \put(150,210){$\bullet$}
     \put(150,270){$\bullet$}
     \put(180,60){$\bullet$}
     \put(180,120){$\bullet$}
     \put(180,180){$\bullet$}
     \put(180,240){$\bullet$}
     \put(210,30){$\bullet$}
     \put(210,90){$\bullet$}
     \put(210,150){$\bullet$}
     \put(210,210){$\bullet$}
     \put(210,270){$\bullet$}
     \put(240,60){$\bullet$}
     \put(240,120){$\bullet$}
     \put(240,180){$\bullet$}
     \put(240,240){$\bullet$}
     \put(270,30){$\bullet$}
     \put(270,90){$\bullet$}
     \put(270,150){$\bullet$}
     \put(270,210){$\bullet$}
     \put(270,270){$\bullet$}
     
    \end{picture}
   \end{center}
   \caption[]{Positions of vortices for $m,\ n=0,\ \pm 1,\ \pm 2,
   \ \pm 3,\ \pm4$ 
   in a constant potential ($a=1$), 
   which are denoted by $\bullet$ 
   and the distance between the neighbouring lines are taken 
   by $\pi/k$.} 
   \label{fig:2.1}
  \end{figure}
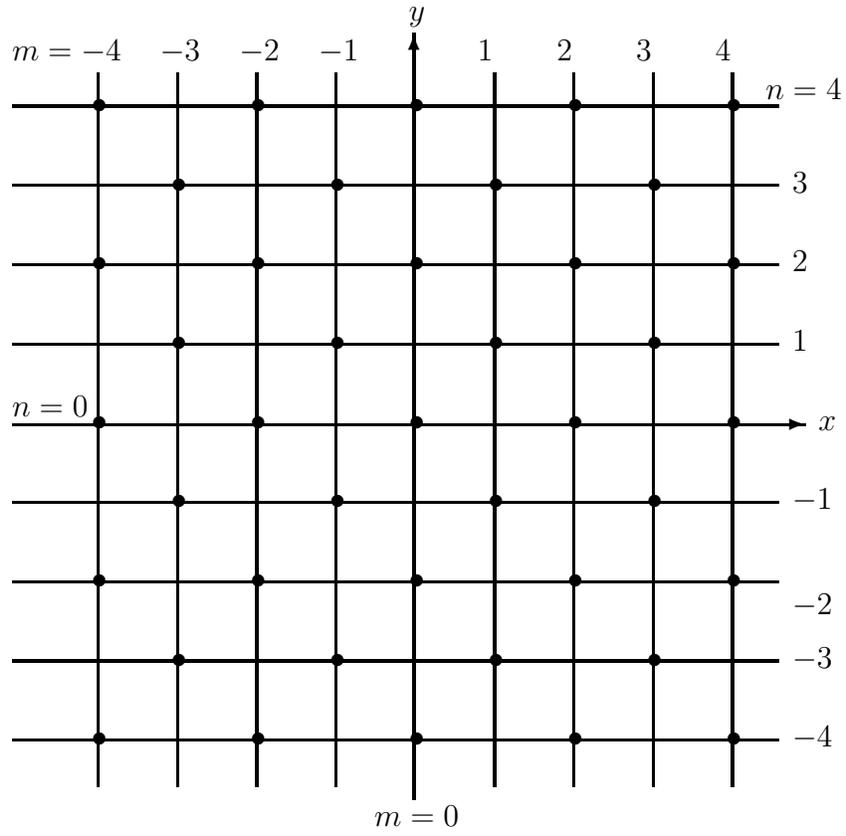 
  
\end{document}